\magnification=1200
\hfuzz=3pt
\hsize=12.5cm
\hoffset=0.42cm
\baselineskip=12pt
\voffset=\baselineskip
\nopagenumbers
\font\title=cmssdc10 at 20pt
\font\smalltitle=cmssdc10 at 14pt

\let\text=\textstyle
\let\display=\displaystyle
\def\RE{\mathop{\Re e}\nolimits}
\def\IM{\mathop{\Im m}\nolimits}

\newfam\bmfam
\font\tenbm=cmmib10

\font\sevenbm=cmmib7

\font\fivebm=cmmib5
\textfont\bmfam=\tenbm
\scriptfont\bmfam=\sevenbm
\scriptscriptfont\bmfam=\fivebm

\def\sectionstyle{\smalltitle}
\newskip\beforesectionskip
\newskip\aftersectionskip
\beforesectionskip=4mm plus 1mm minus 1mm
\aftersectionskip=2mm plus .2mm minus .2mm
\newcount\mysectioncounter
\def\resetsections{\mysectioncounter=0}
\resetsections
\newcount\myeqcounter
\def\mysection#1\par{\par\removelastskip\penalty -250
\vskip\beforesectionskip
\global\advance\mysectioncounter by 1\noindent
\myeqcounter=0{\sectionstyle\the\mysectioncounter.
#1}\par
\nobreak\vskip\aftersectionskip}
\def\myeqno{\global\advance\myeqcounter by 1\eqno{(\the\mysectioncounter.\the\myeqcounter)}}
\def\mydispeqno{\global\advance\myeqcounter by 1\hfill\llap{(\the\mysectioncounter
.\the\myeqcounter)}}
\def\ccomma{\hbox{\raise 1,5pt\hbox{,}}}

\headline={\hfil\tenrm\folio\hfil}
\footline={\hfil}
\baselineskip=12pt
\parindent= 20pt
\centerline{\smalltitle Relaxation times distributions for an anomalously diffusing particle}
\vskip 0.5cm
\centerline{No\"elle POTTIER$^*$}
\medskip
\centerline{\sl Laboratoire Mati\`{e}re et Syst\`{e}mes Complexes (MSC)\/} 
\smallskip
\centerline{\sl UMR 7057, CNRS \& Universit\'e Paris Diderot\/} 
\smallskip
\centerline{\sl F-75205 Paris Cedex 13, France\/}

\vskip 2cm
\noindent
{\smalltitle Abstract}
\bigskip
\noindent
As well-known, the generalized Langevin equation with a memory kernel decreasing at large times as an inverse power law of time describes the motion of an anomalously diffusing particle.\
Here, we focus the interest on some new aspects of the dynamics, successively considering the memory kernel, the particle's mean velocity, and the scattering function.\ All these quantities are studied from a unique angle, namely,  the discussion of the possible existence of a distribution of relaxation times characterizing their time decay.\ Although a very popular concept, a relaxation times distribution cannot be associated with any time-decreasing quantity (from a mathematical point of view, the decay has to be described by a completely monotonic function).\ 

Technically, we use a memory kernel decaying as a Mittag-Leffler function (the Mittag--Leffler functions interpolate between a stretched or compressed exponential behaviour at short times and an inverse power law behaviour at large times).\  We show that, in the case of a subdiffusive motion, relaxation times distributions can be defined for the memory kernel and for the scattering function, but not for the particle's mean velocity.\ The situation is opposite in the superdiffusive case.
\bigskip
\noindent
{\bf Keywords:} Anomalous diffusion; Mittag-Leffler decay; relaxation times distributions.
\vskip 1cm
\parindent=0pt

$^*$ Corresponding author. Tel.: +33 1 57276285; fax: +33 1 57276211
\smallskip
E-mail address: noelle.pottier@univ-paris-diderot.fr
\smallskip
Postal address:  
Laboratoire Mati\`ere et Syst\`emes Complexes (MSC)

B\^atiment Condorcet, CC 7056

F-75205 Paris Cedex 13, France

\vfill
\break
\parskip=0pt
\parindent=20pt
\baselineskip=16pt
\mysection{Introduction}

\noindent
In this paper, we study some new aspects of the dynamics of an anomalously diffusing particle whose motion is described by the generalized Langevin equation:
$$m\dot v(t)+m\int_0^t\gamma(t-t')v(t')\,dt'=F(t),\myeqno$$
where $v(t)$ is the particle velocity, $F(t)$ is the Langevin force acting on the particle, as modelized by a stationary Gaussian random process of zero mean, and $\gamma(t)$ is a dissipative memory kernel.\ Only the knowledge for $t>0$ of this latter function is required in Eq.\ (1.1), but, when necessary, $\gamma(t)$ will be defined for any $t$ as an even function of time.\ As well-known,  for a particle evolving in a bath at thermal equilibrium at a temperature $T$, the consistency of the above model requires that $F(t)$ and $\gamma(t)$ are not independent quantities, but instead are linked by the second fluctuation-dissipation theorem, that is:
$$\langle F(t)F(t')\rangle=mk_BT\gamma(t-t'),\myeqno$$
where $k_B$ is the Boltzmann constant.\ (The symbol $\langle\ldots\rangle$ denotes the average over the realizations of the noise).\ 

In a previous paper [1], we studied this problem in detail in the case of a retarded memory kernel defined by its algebraic decay at large times ($\gamma(t>0)\propto t^{-\delta}$, with $0<\delta<1$ or $1<\delta<2$).\ We showed in particular that the mean-squared displacement of the particle then varies at large times as:
$$\langle\Delta x^2(t)\rangle\sim t^\delta,\myeqno$$
the motion of the particle being thus subdiffusive in the case $0<\delta<1$, and superdiffusive in the case 
$1<\delta<2$.\ The subdiffusive case is especially interesting in relation with experiments in condensed matter physics.\ For instance, the above described model has been used to analyse the subdiffusion process undergone by the distance between a donor and an acceptor of electron transfer within a single protein molecule [2].

In the present paper, following [3], we will use a more general model for the retarded memory kernel, namely, a Mittag-Leffler function of index $\delta$ with the same large time algebraic decay as in [1] and a modified exponential behaviour at short times.\ In this framework, we will focus the interest on some new aspects of the dynamics of the anomalously diffusing particle, successively considering the memory kernel, the particle's mean velocity, and, last, the scattering function.\ All these quantities will be studied from a unique angle, namely, the discussion of the possible existence of a distribution of relaxation times characterizing their time decay.\ The concept of a relaxation times distribution is indeed a very popular one, but the quantity of interest (i.e., for which the existence of a distribution of relaxation times is discussed) needs to be made precise.\ For instance, it is not obvious that, in a given range of values of $\delta$, such a relaxation times distribution exists for each one of the above quoted physical quantities.\ 

The paper is organized as follows.\ In Section 2, we recall the general condition for the existence of a distribution of relaxation times associated with any time-decaying function, and the way this distribution can be obtained.\ In Section 3, we focus the interest on the retarded memory kernel $\gamma(t)$ as described by a Mittag-Leffler function of index $\delta$.\ We propose a microscopic model justifying such a choice.\ To this end, we show how, in the spirit of the Mori-Zwanzig memory function theory, the Mittag-Leffler memory kernel $\gamma(t)$ can itself be viewed as evolving according to a Langevin-like equation associated with a power law second-order memory kernel, a feature which, to the best of our knowledge, does not seem to have been noticed before.\ The possible existence of a relaxation times distribution associated with $\gamma(t)$ is then discussed in detail.\ The particular forms that this distribution (when it exists) takes in the short and long time regimes are discussed.\ In Section~4,
we turn our attention to the particle's mean velocity $\langle v(t)\rangle$.\ We recall its general expression as found in [3], as well as its approximate form in the long time regime as given in [1].\ Then we use this latter expression to discuss the possible existence of a relaxation times distribution associated with $\langle v(t)\rangle$.\ In Section 5, we study the dynamics of the particle over a coarse-grained time scale corresponding to the description of the motion by a non-inertial Langevin equation.\ We introduce the scattering function $F_s(q,t)$, defined as the spatial Fourier transform of the diffusion front, and we discuss the possible existence of a related relaxation times distribution.\ In Section 6, we provide an interpretation for our results, and we summarize our conclusions.
\break

\mysection{General condition for the existence of a distribution of relaxation times}

\noindent
In some instances, the time decay of the physical quantities of interest is described by completely monotonic functions.\ A function $f(t)$ is said to be completely monotonic in the interval $(a,b)$ if  we have [4]:
$${(-1)}^k\,f^{(k)}(t)\geq 0,\qquad a<t<b,\qquad k=0,1,2\ldots\myeqno$$
We will assume $f(t=0)=1$ in what follows.\ 

To begin with, let us recall the Bernstein's theorem, according to which the necessary and sufficient condition for a function $f(t)$ to be completely monotonic in the interval $0\leq t<\infty$ is that it can be written as [4]:
$$f(t)=\int_0^\infty e^{-kt}\,d\alpha(k),\myeqno$$
where $\alpha(k)$ is bounded and non-decreasing, and the above integral converges for $0\leq t<\infty$.\ If $\alpha(k)$ is a function with a continuous first derivative $\mu(k)$, $f(t)$ appears as the direct Laplace transform of $\mu(k)\geq 0$.\ The function $\mu(k)$, normalized such that $\int_0^\infty\mu(k)dk=1$, can be viewed as a probability density of rate constants [5].\ 
\bigskip
\noindent
{\bf 2.1.\ Distribution of relaxation times}

\noindent
From Eq.\ (2.2), we can obtain the distribution of relaxation times associated with the time decay of the completely monotonic function $f(t)$.\ Setting:
$$f(t)=\int_0^\infty P(\tau)\,e^{-{t/\tau}}\,d\tau,\myeqno$$
we get from Eq.\ (2.2) and the definition $\mu(k)={d\alpha(k)/dk}$:
$$P(\tau)={1\over\tau^2}\,\mu\Bigl({1\over\tau}\Bigr)\cdot\myeqno$$
The function $P(\tau)$, normalized such that $\int_0^\infty P(\tau)d\tau=1$, is a  distribution (more precisely, a probability density) of relaxation times.\ Conversely, we have:
$$\mu(k)={1\over k^2}\,P\Bigl({1\over k}\Bigr)\cdot\myeqno$$
\bigskip
\break
\noindent
{\bf 2.2.\ Getting $P(\tau)$ from the Laplace transform of $f(t)$}

\noindent
In some cases, it may be more convenient to start, not from $f(t)$, but from its Laplace transform $\hat f(z)$, as defined by:
$$\hat f(z)=\int_0^\infty f(t)e^{-zt}\,dt.\myeqno$$
The decaying function $f(t)$ can be obtained from $\hat f(z)$ by inverse Laplace transformation:
$$f(t)={1\over 2\pi i}\int_{\rm Br}\hat f(z) e^{zt}\,dz,\myeqno$$
where $\rm Br$ denotes the Bromwich path in the complex $z$-plane, i.e.\ a line ${\RE z}=\sigma$ with $\sigma$ larger than the abscissas of the possibly existing singularities of $\hat f(z)$ and $\IM z$ running from $-\infty$ to $+\infty$.\ For those functions $\hat f(z)$ that have a cut along the negative real axis, it is possible to bend the Bromwich path into a Hankel path, i.e.\ a loop which starts from $-\infty$ along the lower side of the negative real axis, encircles the origin in the positive sense, and ends at $-\infty$ along the upper side of the negative real axis.\ If $\hat f(z)$ has poles in the first Riemann sheet, their contribution has to be taken into account, and it must be added to the contribution of the Hankel path.\ 

This latter contribution to $f(t)$ reads:
$$-{1\over\pi}\int_0^\infty\IM\Bigl\{\hat f(xe^{i\pi})\Bigr\}e^{-xt}\,dx.\myeqno$$
Thus, when no poles are present, the function:
$$-{1\over\pi}\IM\Bigl\{\hat f(ke^{i\pi})\Bigr\}\myeqno$$
can be identified with the probability distribution of rate constants:
$$\mu(k)=-{1\over\pi}\IM\Bigl\{\hat f(ke^{i\pi})\Bigr\},\myeqno$$
from which $P(\tau)$ can be deduced using Eq.\ (2.4).\

\mysection{The Mittag-Leffler retarded memory kernel}

\noindent
Let us now come to the above described purpose, that is, the discussion of some relaxation times distributions intervening in anomalous diffusion.\ Following [3], we will introduce in the generalized Langevin equation (1.1) a Mittag-Leffler retarded memory kernel, namely:
$$\gamma(t)={\gamma_\delta\over  t_0^\delta}\,E_\delta\Bigl[-{\Bigr({t\over t_0}\Bigr)}^\delta\Bigr],\qquad t\geq 0,\myeqno$$
with $0<\delta<2$.\ In Eq.\ (3.1), $ t_0$ acts as a characteristic memory time linked to the particle's environment (thermal bath), and $\gamma_\delta$ is a proportionality coefficient depending on the exponent $\delta$ but independent of time.\ The $E_\alpha(y)$ function denotes the Mittag-Leffler function [6]--[8] defined through the series:
$$E_\alpha(y)=\sum_{n=0}^\infty{y^n\over\Gamma(\alpha n+1)}\ccomma\qquad\alpha>0,\myeqno$$
where $\Gamma$ denotes the Euler Gamma function.\ For $\delta=1$, we just have an ordinary exponential decay as pictured by $\gamma(t)=({\gamma_1/t_0})\exp(-{t/t_0})$.

We assume $\delta\neq 1$ in the rest of this subsection.\ As underlined for instance in [9],  the choice (3.1) is especially convenient, since it allows $\gamma(t)$ to interpolate, from a ``stretched exponential'' (in the case $0<\delta<1$) or a ``compressed exponential'' (in the case $1<\delta<2$) for $t\ll t_0$, to an inverse power law  for $t\gg t_0$.\ More precisely, retaining only the two first terms of the series expansion (3.2), we can approximately write, in the short time regime [9],[10]:
$$\gamma(t)\simeq{\gamma_\delta\over t_0^\delta}\exp\Bigl[-{{({t/ t_0})}^\delta\over\Gamma(\delta+1)}\Bigr]\ccomma\qquad t\ll t_0.\myeqno$$
In the long time regime, we have [6]:
$$\gamma(t)\simeq{\gamma_\delta\over t_0^\delta}\,{1\over\Gamma(1-\delta)}\,{\Bigl({t\over t_0}\Bigr)}^{-\delta}\ccomma\qquad t\gg t_0.\myeqno$$
Note that, in this latter regime, we recover the power law retarded memory kernel previously used in [1].\ Interestingly, the characteristic memory time $t_0$ linked to the bath is in fact not involved in the long time expression for $\gamma(t)$ (Eq.\ (3.4)).\ Setting $\gamma_\delta=\omega_\delta^{2-\delta}$, Eq.\ (3.4) reads:
$$\gamma(t)\simeq{\omega_\delta^2\over\Gamma(1-\delta)}\,{(\omega_\delta t)}^{-\delta},\qquad t\gg t_0 .\myeqno$$
Formula (3.5) shows that the pertinent time scale is then $\omega_\delta^{-1}$, which for consistency we assume to be much larger than $t_0$.\ Summarizing, two different time scales enter into play in the Mittag-Leffler memory kernel (3.1), a short one denoted by $t_0$, and a long one denoted by $\omega_\delta^{-1}$.\ 
\bigskip
\noindent
{\bf 3.1.\ The short time regime}

\noindent
Let us discuss in more detail the short time regime of the Mittag-Leffler function $E_\delta[-{(t/t_0)}^\delta]$, as given by the modified exponential $\exp[-{{({t/ t_0})}^\delta\over\Gamma(\delta+1)}]$ in formula (3.3).\ 

As underlined in [5], the denominations of ``stretched'' and ``compressed'' exponentials are somewhat of a misnomer in the short time regime which precisely interests us now.\ Indeed the so-called stretched exponential $\exp[-{{({t/ t_0})}^\delta\over\Gamma(\delta+1)}]$ in the case $0<\delta<1$ is actually a function whose decay with time is faster than that of the ordinary exponential $\exp[-{{t/ t_0}\over\Gamma(\delta+1)}]$ for  $0<t< t_0$ (but slower afterwards, which is in fact the reason for the denomination of ``stretched'' exponential).\ The initial decay rate of the retarded memory kernel (3.3), which is given by the derivative $[{d\gamma(t)/dt}](t=0)$, is even equal to $-\infty$ in this case.\ 

In the case $1<\delta<2$, the time decay of the so-called compressed exponential $\exp[-{{({t/ t_0})}^\delta\over\Gamma(\delta+1)}]$ is slower than that of the ordinary exponential $\exp[-{{t/ t_0}\over\Gamma(\delta+1)}]$ for $0<t< t_0$ (but faster afterwards, hence  the denomination of ``compressed'' exponential).\ The initial decay rate of the retarded memory kernel vanishes in this case.
\bigskip
\noindent
{\bf 3.2.\ The Laplace transform of the Mittag-Leffler memory kernel}

\noindent
Following the same procedure as in [1], we will solve the generalized Langevin equation (1.1) using Laplace transformation.\ Introducing the Laplace transform $\hat\gamma(z)$ of the retarded memory kernel $\gamma(t)$ defined in the usual way, namely:
$$\hat\gamma(z)=\int_0^\infty\gamma(t)e^{-zt}\,dt,\myeqno$$
we get from Eq.\ (3.1) (as shown for instance in [8]):
$$\hat\gamma(z)={\gamma_\delta\over t_0^\delta}\,{1\over z+ t_0^{-\delta}z^{1-\delta}}\cdot\myeqno$$
For $|z|\ll t_0^{-1}$, we have:
$$\hat\gamma(z)\simeq\gamma_\delta z^{\delta-1},\myeqno$$
that is:
$$\hat\gamma(z)\simeq\omega_\delta^{2-\delta}z^{\delta-1},\myeqno$$
an expression consistent with the choice made in [1], as well as with the large time behaviour of $\gamma(t)$ as given by Eq.\ (3.5).
\bigskip
\noindent
{\bf 3.3.\ A microscopic model for the Mittag-Leffler memory kernel}

\noindent
In the spirit of the Mori-Zwanzig memory function theory [11]--[12], let us assume that  the time evolution of the Langevin force $F(t)$ acting on the particle velocity (a force which is denoted by $F_1(t)$ for more clarity in what follows)  can in turn be described by the Langevin-like equation:
$${\dot F}_1(t)+\int_0^t\phi(t-t')F_1(t')\,dt'=F_2(t),\myeqno$$
where $\phi(t)$ is a second-order memory kernel associated with the time evolution of $F_1(t)$, and $F_2(t)$ acts as a second-order random ``force''.\ Assume in addition that $F_2(t>0)$ is uncorrelated with $F_1(0)$.\ The correlation function $C_1(t)=\langle F_1(t)F_1(0)\rangle$ then obeys the homogeneous evolution equation:
$$\dot C_1(t)+\int_0^t\phi(t-t')C_1(t')\,dt'=0,\qquad t>0.\myeqno$$
The memory kernel $\gamma(t)={C_1(t)/mk_BT}$ of the generalized Langevin equation (1.1) obeys the same evolution equation:
$$\dot\gamma(t)+\int_0^t\phi(t-t')\gamma(t')\,dt'=0,\qquad t>0.\myeqno$$
We want to solve Eq.\ (3.12) for a  given $\gamma(t=0)$.\
Applying Laplace transformation, we get:
$$z\hat\gamma(z)-\gamma(t=0)+\hat\phi(z)\hat\gamma(z)=0.\myeqno$$
We thus recover for $\hat\gamma(z)$ an expression similar to formula (3.7):
$$\hat\gamma(z)={\gamma(t=0)\over z+\hat\phi(z)}\cdot\myeqno$$
More details on the derivation of Eq.\ (3.14) can be found in Appendix B.\ Comparing formulas (3.7) and (3.14), makes it clear that the choice of the Mittag-Leffler memory kernel $\gamma(t)$ as given by Eq.\ (3.1) corresponds to the modelization:
$$\hat\phi(z)= t_0^{-\delta}z^{1-\delta},\myeqno$$
that is, to an inverse power law second-order memory kernel $\phi(t)$:
$$\phi(t)={ t_0^{-2}\over\Gamma(\delta-1)}\,{\Bigl({t\over t_0}\Bigr)}^{\delta-2},\qquad t\geq 0.\myeqno$$
Since $\delta<2$, the function $\phi(t)$ decreases algebraically towards zero at large times ($t\gg t_0$), except for $\delta=1$, in which case this algebraic tail vanishes, leaving us with the delta function behaviour $\phi(t)=({1/t_0})\delta(t)$, in accordance with the formal representation of the Dirac generalized function $\delta(t)={t^{-1}/\Gamma(0)}$, $t\geq 0$ [13].\ Note that only the short time scale $t_0$ is actually involved in the expression for $\phi(t)$ (but not the long time scale $\omega_\delta^{-1}
$).\ We have $\phi(t)<0$ for $0<\delta<1$, and $\phi(t)>0$ for $1<\delta<2$.
\bigskip
\noindent
{\bf 3.4.\ The relaxation times distribution associated with $\gamma(t)$}

\noindent
The question of the existence of a relaxation times distribution associated with a Mittag-Leffler decay has been extensively discussed in the literature [10],[14]--[15].\ We only summarize here the main results.\ Following [8], we write $\gamma(t)$ as the inverse Laplace transform of $\hat\gamma(z)$ as given by formula (3.7):
$$\gamma(t)={\gamma_\delta\over t_0^\delta}\,{1\over 2\pi i}\int_{\rm Br}{1\over z+ t_0^{-\delta}z^{1-\delta}}\,e^{zt}\,dz,\myeqno$$
where $\rm Br$ denotes the Bromwich path in the complex $z$-plane (${\RE z}=\sigma$ with $\sigma\geq t_0^{-1}$).\ The Bromwich path is then bent into the equivalent Hankel path ${\rm Ha}(\varepsilon)$, i.e.\ a loop which starts from $-\infty$ along the lower side of the negative real axis, encircles the origin in the positive sense along a circle of small radius $\varepsilon$, and ends at $-\infty$ along the upper side of the negative real axis.\ In both cases $0<\delta<1$ and $1<\delta<2$, we obtain [8]:
$$\gamma(t)=\gamma_1(t)+\gamma_2(t),\qquad t\geq 0,\myeqno$$
where $\gamma_1(t)$ is the contribution of the Hankel path:
$$\gamma_1(t)={\gamma_\delta\over t_0^\delta}\,{1\over 2\pi i}\int_{\rm Ha(\varepsilon)}{1\over z+ t_0^{-\delta}z^{1-\delta}}\,e^{zt}\,dz,\myeqno$$
and $\gamma_2(t)$ is the contribution of the relevant poles, i.e.\ the poles situated in the first Riemann sheet,  of ${1/(z+t_0^{-\delta}z^{1-\delta})}$.\

In the case $0<\delta<1$, no such poles exist [8], and $\gamma(t)=\gamma_1(t)$ relaxes towards zero without oscillating.\ The result takes a form analogous to Eq.\ (2.2) (except for the fact that we do  not have $\gamma(t=0)=1$, but instead $\gamma(t=0)={\gamma_\delta/t_0^\delta}$):
$$\gamma(t)={\gamma_\delta\over t_0^\delta}\int_0^\infty\mu_\gamma(k)\,e^{-kt}\,dk,\myeqno$$
with the rate constants distribution $\mu_\gamma(k)$ as given by:
$$\mu_\gamma(k)=-{1\over\pi}\,{1\over\gamma(t=0)}\IM\Bigl\{\hat\gamma(ke^{i\pi})\Bigr\}={\sin\delta\pi\over\pi}t_0\,{{(kt_0)}^{\delta-1}\over 1+2{(kt_0)}^\delta\cos\delta\pi+{(kt_0)}^{2\delta}}\cdot\myeqno$$
A distribution of relaxation times can thus be defined in this case (these relaxation times being linked to the thermal bath), which means that we can write:
$$\gamma(t)={\gamma_\delta\over t_0^\delta}\int_0^\infty P_\gamma(\tau)\,e^{-{t/\tau}}\,d\tau,\myeqno$$
with $P_\gamma(\tau)$ a non-negative function normalized such that $\int_0^\infty P_\gamma(\tau)d\tau=1$.\
From Eqs¥ (2.4) and (3.21), we get [8],[10],[13]--[14]:
$$P_\gamma(\tau)={\sin\delta\pi\over\pi}\,{1\over\tau}\,{1\over\display {\Bigl({\tau\over t_0}\Bigr)}^\delta+2\cos\delta\pi+{\Bigl({\tau\over t_0}\Bigr)}^{-\delta}}\cdot\myeqno$$
The distribution $P_\gamma(\tau)$ as given by Eq.\ (3.23) is non-negative and normalized:
$$\int_0^\infty P_\gamma(\tau)\,d\tau=E_\delta(t=0)=1.\myeqno$$
It is widely referred to in the literature as the Cole-Cole distribution of relaxation times [16].\ Its short time and large time limiting forms are discussed in detail in Appendix A.\ The behaviour of $P_\gamma(\tau)$ for $\tau\ll t_0$ is especially interesting.\ We have  an integrable divergence of the distribution of small relaxation times:
$$P_\gamma(\tau)\simeq {\sin\delta\pi\over\pi}\,{1\over t_0}\,{\Bigl({\tau\over t_0}\Bigr)}^{\delta-1},\qquad\tau\ll t_0.\myeqno$$

In the case $1<\delta<2$, the situation is more intricate since $\gamma(t)$ contains in addition an oscillating part.\ It can be written as [8]:
$$\gamma(t)={\gamma_\delta\over t_0^\delta}\,\left\{{2\over\delta}\,\exp\Bigl[{t\over t_0}\cos\Bigl({\pi\over\delta}\Bigr)\Bigr]\cos\Bigl[{t\over t_0}\sin\Bigl({\pi\over\delta}\Bigr)\Bigr]+\int_0^\infty P_\gamma(\tau)\,e^{-{t/\tau}}\,d\tau\right\},\myeqno$$
where the function $P_\gamma(\tau)$, which is still given by Eq.\ (3.23), cannot be given the meaning of a relaxation times distribution.\ We now have:
$$\int_0^\infty P_\gamma(\tau)\,d\tau=E_\delta(t=0)-{2\over\delta}=1-{2\over\delta}\cdot\myeqno$$
Thus, in this range of values of $\delta$, we cannot associate with $\gamma(t)$ a distribution of relaxation times (all the more since now $P_\gamma(\tau)<0$).
\bigskip
\noindent
{\bf 3.5.\ The noise spectral density}

\noindent
Let us come back to the Mittag-Leffler memory kernel $\gamma(t)$ as given by Eq.\ (3.1), and introduce its Fourier-Laplace transform defined as usual for real $\omega$ by:
$$\tilde\gamma(\omega)=\int_0^\infty\gamma(t)e^{i\omega t}\,dt.\myeqno$$
Inversely, we have:
$$\gamma(t)={2\over\pi}\int_0^\infty\RE\tilde\gamma(\omega)\cos\omega t\,d\omega.\myeqno$$
Eq.\ (3.29) displays the fact that it is enough to know $\RE\tilde\gamma(\omega)$ for $\omega>0$ in order to get an integral representation for $\gamma(t)$.\ Using the identity $\tilde\gamma(\omega)=\hat\gamma(z=-i\omega)$, together with the expression (3.7) for $\hat\gamma(z)$, we get (assuming $\omega>0$):
$$\tilde\gamma(\omega)=\gamma_\delta\omega^{\delta-1}\,{\sin({\delta\pi/2})+i\bigl[{(\omega t_0)}^\delta-\cos({\delta\pi/2})\bigr]\over 1+2\cos({\delta\pi/2}){(\omega t_0)}^\delta+{(\omega t_0)}^{2\delta}}\ccomma\qquad\omega>0.\myeqno$$
We have in particular:
$$\RE\tilde\gamma(\omega)=\gamma_\delta\sin({\delta\pi/2})\omega^{\delta-1}\,{1\over 1+2\cos({\delta\pi/2}){(\omega t_0)}^\delta+{(\omega t_0)}^{2\delta}}\ccomma\qquad\omega>0.\myeqno$$

The noise spectral density $\langle{|F(\omega)|}^2\rangle$ defined for any real $\omega$ as the Fourier transform of the correlation function $\langle F(t)F(t')\rangle$, 
reads, making use of the second fluctuation-dissipation theorem (Eq.\ (1.2)):
$$\langle{|F(\omega)|}^2\rangle=2mk_BT\gamma_\delta\sin({\delta\pi/2}){|\omega|}^{\delta-1}\,{1\over 1+2\cos({\delta\pi/2}){(|\omega|t_0)}^\delta+{(|\omega|t_0)}^{2\delta}}\ccomma\myeqno$$
in accordance with the expression provided in [3].\ 

If, instead of the Mittag-Leffler memory kernel, we use a memory kernel $\gamma(t)$ about which we only know that it behaves at times $t\gg t_0$ as an inverse power law of time ($\gamma(t\gg t_0)\propto t^{-\delta}$), we usually write the noise spectral density as [1],[17]:
$$\langle{|F(\omega)|}^2\rangle=2mk_BT\gamma_\delta\sin({\delta\pi/2}){|\omega|}^{\delta-1}\,f_c(|\omega|t_0),\myeqno$$
with $f_c$ a high-frequency cut-off function of typical width $t_0^{-1}$.\  For $\delta\neq 1$, we have at hand a non-Ohmic dissipation model [17] (more precisely, sub-Ohmic for $0<\delta<1$, and super-Ohmic for $1<\delta<2$).\ The spectral density of the coupling, defined for positive $\omega$ by $J(\omega)=m\RE\tilde\gamma(\omega)$, is given by:
$$J(\omega)=m\gamma_\delta\sin({\delta\pi/2})\omega^\delta f_c(\omega t_0),\qquad\omega>0.\myeqno$$
These results are in accordance with those obtained in [18], where it has been shown, in the framework of the Kac-Zwanzig model of a particle interacting through springs with a heat bath made of a collection of harmonic oscillators [19], how the distributions of frequencies, masses, and spring constants, determine the parameters of the generalized Langevin equation obeyed by the distinguished particle in the thermodynamic limit.\ The analysis carried out in [18] has been applied to the case of a memory kernel decreasing as an inverse power law of time.\ 

Let us come back to the full formula (3.32) for the noise spectral density.\ As put forward in [3], the choice of the Mittag-Leffler memory kernel allows us to make precise the cut-off function, which depends on the exponent $\delta$:
$$f_c(|\omega|t _0)={1\over 1+2\cos({\delta\pi/2}){(|\omega|t_0)}^\delta+{(|\omega|t_0)}^{2\delta}}\cdot\myeqno$$
Note that, for $\delta=1$, formula (3.35) for the cut-off function reduces to the Drude expression:
$$f_c(|\omega|t_0)={1\over 1+{(|\omega|t_0)}^2}\cdot\myeqno$$
Interestingly, instead of being introduced by hand (usual choices being an exponential or a Lorentzian cut-off function [17]) the $\delta$-dependent cut-off function comes here into play in a natural way in the framework of the Mori-Zwanzig formalism (see Appendix B for more details).

\mysection{The particle's mean velocity}

\noindent
The solution $\langle v(t)\rangle$ of the generalized Langevin equation (1.1) with $\gamma(t)$ the Mittag-Leffler memory kernel as studied in Section 3 has first been obtained in [3] as the inverse Laplace integral:
$$\langle v(t)\rangle=v(t=0){1\over 2\pi i}\int_{\rm Br}{1\over z+\hat\gamma(z)}\,e^{zt}\,dz,\myeqno$$
with $\hat\gamma(z)$ as given by Eq.\ (3.7).\ To get an explicit formula, we write:
$${1\over z+\hat\gamma(z)}={1\over z}\sum_{k=0}^\infty{(-1)}^k{{\bigl[\hat\gamma(z)\bigr]}^k\over z^k}\cdot\myeqno$$
The expression in the r.h.s.\ of Eq.\ (4.2) can conveniently be rewritten as:
$$\sum_{k=0}^\infty{{(-1)}^k\over k!}\,{k!\over z^{k+1}}{\Bigl({\gamma_\delta\over t_0^\delta}\Bigr)}^k\,{z+t_0^{-\delta} z^{1-\delta}\over{(z+t_0^{-\delta} z^{1-\delta})}^{k+1}}\ccomma\myeqno$$
that is:
$$\sum_{k=0}^\infty{{(-1)}^k\over k!}\,{k!\over z^k}{\Bigl({\gamma_\delta\over t_0^\delta}\Bigr)}^k\,{1\over{(z+t_0^{-\delta} z^{1-\delta})}^{k+1}}+\sum_{k=0}^\infty{{(-1)}^k\over k!}\,{k!\over z^{k+\delta}}{\Bigl({\gamma_\delta\over t_0^\delta}\Bigr)}^k\,{t_0^{-\delta}\over{(z+t_0^{-\delta} z^{1-\delta})}^{k+1}}\cdot\myeqno$$
The result of the Laplace inversion takes the form of the sum of two infinite series of derivatives of generalized Mittag-Leffler functions (see for instance the formulas provided in [7]):
$$\displaylines{\quad\langle v(t)\rangle=v(t=0)\,\Biggl\{\sum_{k=0}^\infty{{(-1)}^k\over k!}\,{\Bigl({\gamma_\delta\over t_0^\delta}\Bigr)}^k t^{2k}E_{\delta,1+(2-\delta)k}^{(k)}\Bigl[-{\Bigl({t\over t_0}\Bigr)}^\delta\Bigr]\hfill\cr\noalign{\vskip 5pt}
\hfill{}+{\Bigl({t\over t_0}\Bigr)}^\delta\sum_{k=0}^\infty{{(-1)}^k\over k!}\,{\Bigl({\gamma_\delta\over t_0^\delta}\Bigr)}^k t^{2k}E_{\delta,1+\delta+(2-\delta)k}^{(k)}\Bigl[-{\Bigl({t\over t_0}\Bigr)}^\delta\Bigr]\Biggr\}.\quad\cr\mydispeqno}$$
The generalized Mittag-Leffler function $E_{\alpha,\beta}(y)$ is defined through the series [6]--[7]:
$$E_{\alpha,\beta}(y)=\sum_{n=0}^\infty{y^n\over\Gamma(\alpha n+\beta)}\ccomma\qquad\alpha>0\qquad\beta>0,\myeqno$$
and $E_{\alpha,\beta}^{(k)}(y)$ represents its $k$th-order derivative with respect to its argument $y$:
$$E_{\alpha,\beta}^{(k)}(y)={d^k\over dy^k} E_{\alpha,\beta}(y).\myeqno$$

The main question we now want to address is that of the possible existence of a distribution of relaxation times associated with the particle's mean velocity.\ This discussion would be tricky to achieve on the basis of the full exact formula (4.5).\ This is why, from now on, we will use a simpler expression for $\langle v(t)\rangle$, valid for times $t\gg t_0$.
\bigskip
\noindent
{\bf 4.1.\ The average velocity at times $t\gg t_0$}

\noindent
Let us thus take for $\hat\gamma(z)$  its approximate expression (3.8) valid for $|z|\ll t_0^{-1}$, which amounts to use, in the time domain, the expression (3.5) for $\gamma(t)$, valid for times $t\gg t_0$.\ 
We then write accordingly (with obvious notations) [1]:
$$\langle\hat v(z)\rangle\simeq{v(t=0)\over z+\omega_\delta^{2-\delta}z^{\delta-1}}\ccomma\qquad|z|\ll t_0^{-1},\myeqno$$
which gives:
$$\langle v(t)\rangle\simeq v(t=0)E_{2-\delta}\bigl[-{(\omega_\delta t)}^{2-\delta}\bigr],\qquad t\gg t_0.\myeqno$$
As shown in [3], this behaviour can also be retrieved from the general expression (4.5).\ Replace, in the two infinite series in the r.h.s.\ of Eq.\ (4.5), each generalized Mittag-Leffler function by its asymptotic expression for $t\gg t_0$, using the asymptotic formula [7]:
$$E_{\alpha,\beta}(-y)\simeq{1\over y\Gamma(\beta-\alpha)}\ccomma\qquad y>0.\myeqno$$
The leading order terms, which come from the second series in formula (4.5), can be resummed.\ The pertinent time scale is now $\omega_\delta^{-1}$ (the shorter time scale $t_0$ being no more involved).\ We recover in this way formula (4.9).
\bigskip
\noindent
{\bf 4.2. Connection with the fractional Langevin equation}

\noindent
To shed some more light on the above results, let us remark that, using for $\gamma(t)$ the approximate expression (3.5), amounts, in the case $0<\delta<1$, to writing the generalized Langevin equation (1.1) as a fractional Langevin equation, namely:
$$m\dot v(t)+m{\gamma_\delta\over\Gamma(1-\delta)}\int_0^t{(t-t')}^{-\delta}v(t')\,dt'=F(t),\qquad 0<\delta<1.\myeqno$$
In the range $1<\delta<2$, we write instead, in order to avoid divergencies:
$$m\dot v(t)+m\gamma_\delta{v(t=0)t^{1-\delta}\over\Gamma(2-\delta)}+m{\gamma_\delta\over\Gamma(2-\delta)}\int_0^t{(t-t')}^{-\delta+1}\dot v(t')\,dt'=F(t),\quad 1<\delta<2.\myeqno$$
Accordingly, the average particle's velocity as given by Eq.\ (4.9) coincides as it should with the (averaged) solution of Eq.\ (4.11) (for $0<\delta<1$), or of Eq.\ (4.12) (for $1<\delta<2$). 

In addition, let us note that formulas  (4.8) and (4.9) for the average particle's velocity have first been provided in [20] for $0<\delta<1$ as the force-free solution of a fractional Fokker-Planck equation, as well as in [21] for $1<\delta<2$ as the force-free solution of a fractional Kramers equation, this latter equation describing both the velocity and coordinate of a particle exhibiting anomalous diffusion in an external force field (see also [9]).
\bigskip
\noindent
{\bf 4.3.\ The noise spectral density}

\noindent
Let us here quote for completeness the corresponding expressions for the noise spectral density and the spectral density of the coupling.\ In the angular frequency range $|\omega|\ll t_0^{-1}$, formula (3.31) for $\RE\tilde\gamma(\omega)$ reduces to the expression used in [1], namely:
$$\RE\tilde\gamma(\omega)\simeq\gamma_\delta\sin({\delta\pi/2}){|\omega|}^{\delta-1}.\myeqno$$
The corresponding noise spectral density (Eq.\ (3.33)) now reads:
$$\langle{|F(\omega)|}^2\rangle\simeq2mk_BT\gamma_\delta\sin({\delta\pi/2}){|\omega|}^{\delta-1}.\myeqno$$
As for the spectral density of the coupling (Eq.\ (3.34)), it reduces to:
$$J(\omega)\simeq m\gamma_\delta\sin({\delta\pi/2})\omega^\delta,\qquad\omega>0.\myeqno$$
Formulas (4.13), (4.14), and (4.15) are deduced from their counterparts valid for any $\omega$ (respectively, formulas (3.31), (3.33), and (3.34)) by making $f_c(|\omega|t_0)\simeq 1$ (indeed the cut-off function does not play any role in the angular frequency range $|\omega|t_0\ll 1$).\
\bigskip
\noindent
{\bf 4.4.\ Duality symmetry}

\noindent
Let us compare formula (3.14) for $\hat\gamma(z)$ (with $\hat\phi(z)$ as given by Eq.\ (3.15)), with the approximate formula (4.8) for $\langle\hat v(z)\rangle$.\ Both expressions have the same structure.\ They can even be identified, provided that one makes the change of indexes:
$$\delta\longleftrightarrow 2-\delta,\myeqno$$
together with the change of time scales:
$$t_0\longleftrightarrow\omega_\delta^{-1}.\myeqno$$

This dynamical mapping between the Laplace transforms of the memory kernel, on the one hand, and of the particle's mean velocity, on the other hand, transforms the parameter region $0<\delta<1$ into the region $1<\delta<2$ (and vice-versa), with important consequences about the existence of relaxation times distributions for either quantities.\ 
\bigskip
\noindent
{\bf 4.5.\ The relaxation times distribution associated with $\langle v(t)\rangle$}

\noindent
We will take advantage of the duality symmetry to discuss the existence of a relaxation times distribution associated with the particle's mean velocity as given by the approximate formula (4.9).\ In both cases $0<\delta<1$ and $1<\delta<2$, we write:
$$\langle v(t)\rangle=\langle v_1(t)\rangle+\langle v_2(t)\rangle,\qquad t\geq 0,\myeqno$$
where:
$$\langle v_1(t)\rangle=v(t=0){1\over 2\pi i}\int_{\rm Ha(\varepsilon)}{1\over z+\omega_\delta^{2-\delta}z^{\delta-1}}e^{zt}\,dz,\myeqno$$
and $\langle v_2(t)\rangle$ is the contribution of the relevant poles of ${1/(z+\omega_\delta^{2-\delta}z^{\delta-1})}$, if any.

In the case $1<\delta<2$, no such poles exist, and the mean velocity relaxes towards zero without oscillating.\ We have:
$$\langle v(t)\rangle=v(t=0)\int_0^\infty P_{\langle v\rangle}(\tau)\,e^{-{t/\tau}}\,d\tau,\myeqno$$
with the relaxation times distribution $P_{\langle v\rangle}(\tau)$ as given by:
$$P_{\langle v\rangle}(\tau)={-\sin\delta\pi\over\pi}\,{1\over\tau}\,{1\over\display {\bigl(\omega_\delta\tau\bigr)}^{2-\delta}+2\cos\delta\pi+{\bigl(\omega_\delta\tau\bigr)}^{\delta-2}}\cdot\myeqno$$

In the case $0<\delta<1$, the particle's mean velocity contains in addition an oscillating part:
$$\displaylines{\quad\langle v(t)\rangle=v(t=0)\,\Biggl\{{2\over 2-\delta}\,e^{\omega_\delta t\cos[{\pi/(2-\delta})]}\cos\Bigl[\omega_\delta t\sin\Bigl({\pi\over 2-\delta}\Bigr)\Bigr]\hfill\cr
\hfill{}+\int_0^\infty P_{\langle v\rangle}(\tau)\,e^{-{t/\tau}}\,d\tau\Biggr\},\quad\cr\mydispeqno}$$
where $P_{\langle v\rangle}(\tau)$, although still given by Eq.\ (4.21), cannot obviously be interpreted as a relaxation times distribution.\ We easily check that, when $\delta\to 1$, we have $\omega_\delta\to\gamma_1$, and $P_{\langle v\rangle}(\tau)\to-\delta(\tau-\gamma_1^{-1})$.\ Formula (4.22) thus consistently yields in this limit the standard behaviour corresponding to a non-retarded Langevin equation with friction coefficient $\gamma_1$:
$$\langle v(t)\rangle\to v(t=0)\,e^{-\gamma_1t}.\myeqno$$
In the opposite limit $\delta\to 0$, setting $\omega_\delta\to\omega_0$, we have an undamped oscillation:
$$\langle v(t)\rangle\to v(t=0)\cos\omega_0t.\myeqno$$
A physical explanation for the presence of an oscillating term in the r.h.s.\ of Eq.\ (4.22) can be provided in terms of the cage effect.\ As explained in [22], for small $\delta$ the friction force induced by the environment is not just slowing down the particle, but also causes the particle a rattling motion.\ The generalized Langevin equation can be written, for $0<\delta<1$, as the fractional Langevin equation (4.11), that is:
$$m\dot v(t)+m{\omega_\delta^2\over\Gamma(1-\delta)}\int_0^t{[\omega_\delta(t-t')]}^{-\delta}v(t')\,dt'=F(t).\myeqno$$
In the limit $\delta\to 0$, Eq.\ (4.25) simply reads:
$$m\dot v(t)+m\omega_0^2[x-x(t=0)]=F(t).\myeqno$$
This shows that a purely oscillating behaviour is expected in this limit.\ This oscillation can be explained by the rattling motion of the particle of interest in the cage formed by the surrounding particles.
\bigskip
\noindent
{\bf 4.6.\ Discussion}

\noindent
Let us comment the above found results concerning the fact that it is not possible to define  a relaxation times distribution for the particle's average velocity in the range $0<\delta<1$.\

Consider an assembly of independent particles obeying the usual simple Lange\-vin equation:
$$m\dot{v}(t)+m{v(t)\over\tau}=F(t),\myeqno$$
where $F(t)$ is the Langevin force and the relaxation time $\tau$ is random.\ For any given $\tau$, the average velocity of an individual particle varies like $v(t=0)\exp(-{t/\tau})$.\ An obvious question is whether it exists a relaxation times distribution function $P_{\langle v\rangle}(\tau)$ such that the average velocity as defined by:
$$\langle v(t)\rangle=v(t=0)\int_0^\infty P_{\langle v\rangle}(\tau)\,e^{-{t/\tau}}\,d\tau,\myeqno$$
effectively obeys the (averaged) fractional Langevin equation, namely:
$$m\langle\dot v(t)\rangle+m{\gamma_\delta\over\Gamma(1-\delta)}\int_0^t{(t-t')}^{-\delta}\langle v(t')\rangle\,dt'=0,\qquad 0<\delta<1.\myeqno$$
It ensues from the results of Subsection 4.5 that such a distribution function cannot be defined.\ In other words, it does not exist a relaxation times distribution which would map the simple Langevin equation onto the fractional one.

\mysection{The scattering function}

\noindent
We continue in this section, and in the following of this paper as well, to study the various functions of time which enter into play at times $t\gg t_0$ only.\ This latter time scale will thus no more appear in what follows.
\bigskip
\noindent
{\bf 5.1.\ The mean-squared displacement}

\noindent
Let us introduce the particle displacement $x(t)=\int_0^t v(t')\,dt'$.\ The two-time displacement correlation function can be obtained from the velocity correlation function as the double integral:
$$\langle x(t)x(t')\rangle=\int_0^t\int_0^{t'}\langle v(t_1)v(t_2)\rangle\,dt_1dt_2.\myeqno$$
On the basis of the first fluctuation-dissipation theorem (see formula (B.6)) together with formula (4.9) for the particle's mean velocity, the velocity correlation function is given by [1],[9]:
$$\langle v(t)v(t')\rangle={k_BT\over m}E_{2-\delta}\bigl[-{(\omega_\delta|t-t'|)}^{2-\delta}\bigr].\myeqno$$
To obtain $\langle x(t)x(t')\rangle$, we have to use the integration formulas:
$$\int_0^tE_{2-\delta}\bigl[-{(\omega_\delta t_1\bigr)}^{2-\delta}\bigr]\,dt_1=t\,E_{2-\delta,2}\bigl[-{(\omega_\delta t)}^{2-\delta}\bigr],\myeqno$$
and:
$$\int_0^tdt_1\int_0^{t_1}E_{2-\delta}\bigl[-{(\omega_\delta t_2\bigr)}^{2-\delta}\bigr]\,dt_2=t^2\,E_{2-\delta,3}\bigl[-{(\omega_\delta t)}^{2-\delta}\bigr].\myeqno$$
Making $t'=t$ in Eq.\ (5.1), gives the expression for the mean-squared displacement $\langle\Delta x^2(t)\rangle=\langle x^2(t)\rangle$ [1],[9],[21]:
$$\langle\Delta x^2(t)\rangle=2{k_BT\over m}\,t^2\,E_{2-\delta,3}\bigl[-{(\omega_\delta t)}^{2-\delta}\bigr].\myeqno$$

In the case $0<\delta<1$, the expression for $\langle\Delta x^2(t)\rangle$ contains an oscillating part.\ The general formula being a bit lengthy, we shall only quote its two limiting forms for $\delta\to 1$ and $\delta\to 0$.\ First, we can check that, when $\delta\to 1$, we recover the standard Brownian motion result:
$$\langle\Delta x^2(t)\rangle\to 2{k_BT\over m}\Bigl({t\over\gamma_1}-{1-e^{-\gamma_1t}\over\gamma_1^2}\Bigr)\cdot\myeqno$$
In the opposite limit $\delta\to 0$, we get the undamped oscillator formula:
$$\langle\Delta x^2(t)\rangle=2{k_BT\over m\omega_0^2}\,(1-\cos\omega_0t).\myeqno$$

In the case $1<\delta<2$ in which the motion is superdiffusive, we can write the mean-squared displacement  in terms of the relaxation times distribution function $P_{\langle v\rangle}(\tau)$ as:
$$\langle\Delta x^2(t)\rangle=2{k_BT\over m}\int_0^\infty P_{\langle v\rangle}(\tau)\bigl[t-\tau(1-e^{-{t/\tau}})\bigr]\tau\,d\tau.\myeqno$$
\bigskip
\noindent
{\bf 5.2.\ Large time behaviour of the mean-squared displacement: the noninertial regime}

\noindent
We mean here by ``large times'' the times for which we have $\omega_\delta t\gg 1$.\ The generalized Mittag-Leffler function in Eq.\ (5.5) can then be replaced by its leading order term as given by the asymptotic formula (4.10).\ We get [1]:
$$\langle\Delta x^2(t)\rangle\simeq 2{k_BT\over m\omega_\delta^2}\,{{(\omega_\delta t)}^\delta\over\Gamma(\delta+1)}\ccomma\qquad\omega_\delta t\gg 1.\myeqno$$
Formula (5.9) is valid for any $\delta$ ($0<\delta<2$).\ Note that, when $\omega_\delta t\gg 1$, the oscillating contribution which exists at shorter times in the range $0<\delta<1$ does not appear any more.\ 

Interestingly, the result (5.9) is fully consistent with the expression for the velocity correlation function obtained by considering, instead of the full generalized Langevin equation (1.1), the corresponding noninertial equation of motion, i.e.:
$$m\int_0^t\gamma(t-t')v(t')\,dt'=F(t).\myeqno$$
Applying Laplace transformation to Eq.\ (5.10), we get:
$$m\hat\gamma(z)\hat v(z)=\hat F(z),\myeqno$$
hence the expression for the double Laplace transform of the velocity correlation function:
$$\langle v(z)\hat v(z')\rangle={1\over m^2}{\langle\hat F(z)\hat F(z')\rangle\over\hat\gamma(z)\hat\gamma(z')}\cdot\myeqno$$
From the second fluctuation-dissipation theorem, we have [1]:
$$\langle\hat F(z)\hat F(z')\rangle=mk_BT{\hat\gamma(z)+\hat\gamma(z')\over z+z'}\ccomma\myeqno$$
and thus:
$$\langle\hat v(z)\hat v(z')\rangle={k_BT\over m}{1\over z+z'}\Bigl({1\over\hat\gamma(z)}+{1\over\hat\gamma(z')}\Bigr)\cdot\myeqno$$
Taking for $\hat\gamma(z)$ its expression (3.9) valid for $|z|\ll t_0^{-1}$, we get:
$$\langle\hat v(z)\hat v(z')\rangle={k_BT\over m}{\omega_\delta^{\delta-2}\over z+z'}\,(z^{1-\delta}+z'^{1-\delta}).\myeqno$$
Neglecting the inertial term thus yields for the velocity correlation function the expression:
$$\langle v(t)v\rangle={k_BT\over m}\,{{(\omega_\delta t)}^{\delta-2}\over\Gamma(\delta-1)}\cdot\myeqno$$
This latter formula is identical to the one which can be deduced by derivating twice with respect to time the large time expression (5.9) for $\langle\Delta x^2(t)\rangle$.\ This fact is in accordance with the general view according to which considering the noninertial equation of motion is tantamount to studying the motion of the particle over a coarse-grained time scale (here, $\omega_\delta^{-1}$).
\bigskip
\noindent
{\bf 5.3.\ The scattering function} 

\noindent
Let us introduce the scattering function $F_s(q,t)$, as defined by [23]:
$$F_s(q,t)=\int G_s(x,t)\,e^{iqx}\,dx,\myeqno$$
with $G_s(x,t)$ the diffusion profile (if the present one-particle model is considered as picturing the diffusion of an ensemble of non-interacting particles, the function $F_s(q,t)$ describes the relaxation of a density fluctuation of wavevector $q$).\ Since the noise has been assumed to be Gaussian, the diffusion profile is a Gaussian function of width $\langle\Delta x^2(t)\rangle$:
$$G_s(x,t)={[2\pi\langle\Delta x^2(t)\rangle]}^{-{1/2}}\exp\Bigl[-{x^2\over 2\langle\Delta x^2(t)\rangle}\Bigr]\cdot\myeqno$$
Performing the Fourier integration in Eq.\ (5.17), we get:
$$F_s(q,t)=\exp\Bigl[-q^2{\langle\Delta x^2(t)\rangle\over 2}\Bigr]\cdot\myeqno$$

Using Eq.\  (5.5), which gives the expression for the mean-squared displacement in terms of a generalized Mittag-Leffler function, would lead to a fairly involved formula for $F_s(q,t)$.\ Let us instead simply use for $\langle\Delta x^2(t)\rangle$ its expression (5.9) valid for $\omega_\delta t\gg 1$ (non-inertial regime).\ The scattering function is then simply a modified exponential function [24]:
$$F_s(q,t)\simeq\exp\left[-q^2{k_BT\over m\omega_\delta^2}{{(\omega_\delta t)}^\delta\over\Gamma(\delta+1)}\right]\ccomma\qquad\omega_\delta t\gg 1.\myeqno$$
\bigskip
\noindent
{\bf 5.4.\ The relaxation times distribution associated with $F_s(q,t)$}

In the range $0<\delta<1$ which corresponds to subdiffusion, the scattering function as given by formula (5.20) is a stretched exponential, whereas, in the range $1<\delta<2$ in which the motion is superdiffusive, it is a compressed exponential.\ Accordingly, a relaxation times distribution can be associated with the scattering function in the range $0<\delta<1$, but not in the range $1<\delta<2$.\ 

Introducing the time scale $t_q$ as defined by:
$${1\over t_q^\delta}=q^2{k_BT\over m\omega_\delta^2}\omega_\delta^\delta,\myeqno$$
we can rewrite Eq.\ (5.20) as:
$$F_s(q,t)\simeq\exp\Bigl[-{{({t/ t_q})}^\delta\over\Gamma(\delta+1)}\Bigr]\ccomma\qquad \omega_\delta t\gg 1.\myeqno$$
Formula (5.22) for $F_s(q,t)$ has the same form as the approximate formula (3.3) for $\gamma(t)$ valid for $t\ll t_0$.\ Both formulas can be identified (up to a proportionality factor), provided that one makes the change of time scales:
$$t_0\longleftrightarrow t_q,\myeqno$$
the parameter $\delta$ remaining unchanged.\

In the range $0<\delta<1$, introducing the rates distribution $\mu_{F_s}(k)$, we write:
$$F_s(q,t)=\int_0^\infty\mu_{F_s}(k)\,e^{-kt}\,dk,\myeqno$$
that is, in terms of the relaxation times distribution $P_{F_s}(\tau)$:
$$F_s(q,t)=\int_0^\infty P_{F_s}(\tau)\,e^{-{t/\tau}}\,d\tau.\myeqno$$
The relaxation rates distribution $\mu_{F_s}(k)$ in Eq. (5.24), as given by:
$$\mu_{F_s}(k)={1\over 2\pi i}\int_{\rm Br}\exp\Bigl[-{{(z/t_q)}^\delta\over\Gamma(\delta+1)}\Bigr]\,e^{kz}\,dz,\myeqno$$
identifies with the one-sided stable L\'evy law $L^{^C}_{\delta,1}(k)$ as defined by [25]:
$$L^{^C}_{\delta,1}(k)={1\over 2\pi i}\int_{\rm Br}e^{-Cz^\delta}\,e^{kz}\,dz,\myeqno$$
provided that one sets $C={t_q^{-\delta}/\Gamma(\delta+1)}$.\ This property deserves to be mentioned since it relates the fractional Langevin equation to the generalized central limit theorem [25].

Carrying out a discussion similar to the one in Appendix A, we write the relaxation rates distribution $\mu_{F_s}(k)$ as:
$$\mu_{F_s}(k)={1\over\pi}\,t_q\int_0^\infty\exp\Bigl[-{x^\delta\cos\delta\pi\over\Gamma(\delta+1)}\Bigr]\,\sin\Bigl[{x^\delta\sin\delta\pi\over\Gamma(\delta+1)}\Bigr]\,e^{-kt_qx}\,dx.\myeqno$$
As for the relaxation times distribution $P_{F_s}(\tau)$, it reads:
$$P_{F_s}(\tau)={1\over\pi}\,{t_q\over\tau^2}\int_0^\infty\exp\Bigl[-{x^\delta\cos\delta\pi\over\Gamma(\delta+1)}\Bigr]\,\sin\Bigl[{x^\delta\sin\delta\pi\over\Gamma(\delta+1)}\Bigr]\,e^{-{xt_q/\tau}}\,dx.\myeqno$$
For $\tau\ll t_q$, we simply have:
$$P_{F_s}(\tau)\simeq{\sin\delta\pi\over\pi}\,{1\over\tau}\,\bigl({\tau\over t_q}\Bigr)^\delta.\myeqno$$

In the range $1<\delta<2$, $F_s(q,t)$ is a compressed exponential with which no relaxation times distribution can be associated.\ Finally, let us note for completeness that, in the limiting case $\delta=1$, we just recover an exponentially decaying scattering function:
$$F_s(q,t)\simeq\exp(-{t/t_q}),\qquad\gamma_1t\gg 1,\myeqno$$
with:
$${1\over t_q}=q^2{k_BT\over m\gamma_1}\cdot\myeqno$$

\mysection{Conclusion}

\noindent
In this paper, we studied the dynamics of an anomalously diffusing particle whose motion is described by a generalized Langevin equation and whose mean-squared displacement grows at large time like $\langle\Delta x^2(t)\rangle\sim t^\delta$, with $0<\delta<1$ (subdiffusion) or $1<\delta<2$ (superdiffusion).\  We successively focused our interest on three time decaying quantities, namely, the memory kernel of the generalized Langevin equation, the particle's mean velocity, and the scattering function.\

Using for the memory kernel $\gamma(t)$ a Mittag-Leffler function of index $\delta$ and characteristic time scale $t_0$, we showed that its decay can be associated with a relaxation times distribution in the range $0<\delta<1$.\ We were able to make precise this distribution at any time $t$, as well as to check its consistency with the relaxation times distributions associated with the limiting forms of the Mittag-Leffler function for $t\ll t_0$ (a stretched exponential) and for $t\gg t_0$ (an inverse power law of time).\ Interestingly, the distribution of small relaxation times $\tau\ll t_0$ was shown to exhibit an integrable divergence $\propto\tau^{\delta-1}$.\ As stated in [24], such a divergence implies a profusion of short times scales in the Mittag-Leffler decay of index $\delta$ in the range $0<\delta<1$.\ As for the distribution of large relaxation times, it was shown to decay like an inverse power law $\propto\tau^{-\delta}$, which displays the fact that there exist also abundant long time scales in the Mittag-Leffler decay.

We then turned to the particle's mean velocity $\langle v(t)\rangle$.\ Restricting the study to times $t\gg t_0$, we showed that $\langle v(t)\rangle$ is given by a Mittag-Leffler function of index $2-\delta$ and characteristic time scale $\omega_\delta^{-1}\gg t_0$, and thus that it can be associated with a relaxation times distribution in the range $1<\delta<2$.\ More precisely, the dynamical mapping which allows us to deduce the behaviour of $\langle v(t)\rangle$ from that of $\gamma(t)$ was shown to correspond to the change of indexes $\delta\longleftrightarrow 2-\delta$, together with the change of time scales $t_0\longleftrightarrow\omega_\delta^{-1}$.\ The divergence of the distribution of small relaxation times of $\gamma(t)$, all the more pronounced as $\delta$ approaches zero, induces for $0<t<t_0$ an efficient decrease of the dissipative kernel, resulting in turn in a weak damping of the particle's mean velocity, which is actually described by an oscillating function in the range $0<\delta<1$.\ Conversely, for $1<\delta<2$, the decrease of $\gamma(t)$ for $0<t<t_0$ is relatively weak, which in turn implies an efficient decrease of the particle's mean velocity.

Finally, we turned our interest on the scattering function $F_s(q,t)$, whose behaviour is directly linked to that of the mean-squared displacement.\ It results from the very definition of the scattering function that this quantity can be associated with a relaxation time distribution in the subdiffusive range $0<\delta<1$, but not in the superdiffusive range $1<\delta<2$. 
\vskip 1cm

\noindent
{\smalltitle Acknowledgement}
\bigskip
\noindent
We thank the referees for helpful suggestions and comments.

\vfill
\break
\noindent
{\smalltitle Appendix A}
\bigskip
\noindent
Let us check the consistency, in the case $0<\delta<1$, between the relaxation times distributions obtained for the Mittag-Leffler function $E_\delta[-{(t/t_0)}^\delta]$, on the one hand, and for the stretched exponential function and the algebraically decreasing function respectively describing its short time and long time behaviours, on the other hand.
\bigskip
\noindent
{\bf A.1.\ The relaxation times distribution associated with the stretched exponential $\exp[-{{({t/ t_0})}^\delta\over\Gamma(\delta+1)}]$ ($0<\delta<1$)}

\noindent
Let us here discuss the existence of a distribution of relaxation times for the modified exponential function $\exp[-{{({t/ t_0})}^\delta\over\Gamma(\delta+1)}]$ which pictures the short time behaviour of the Mittag-Leffler function $E_\delta[-{(t/t_0)}^\delta]$.

The stretched exponential $\exp[-{{({t/ t_0})}^\delta\over\Gamma(\delta+1)}]$ with $0<\delta<1$ is completely monotonic  (it is often referred to as the Kohlrausch-Williams-Watt  (KWW) relaxation law [26]), but this is not the case for the compressed exponential  $\exp[-{{({t/ t_0})}^\delta\over\Gamma(\delta+1)}]$ with $1<\delta<2$.\ It is therefore solely in the range $0<\delta<1$ that a relaxation times distribution, denoted by $P_{\rm KWW}(\tau)$,  can be associated with the modified exponential $\exp[-{{({t/ t_0})}^\delta\over\Gamma(\delta+1)}]$.\ We set:
$$\exp\Bigl[-{{({t/ t_0})}^\delta\over\Gamma(\delta+1)}\Bigr]=\int_0^\infty\mu_{\rm KWW}(k)\,e^{-kt}\,dk.\eqno({\rm A}.1)$$
In Eq. (A.1), the relaxation rates distribution $\mu_{\rm KWW}(k)$ as given by:
$$\mu_{\rm KWW}(k)={1\over 2\pi i}\int_{\rm Br}\exp\Bigl[-{{(z/t_0)}^\delta\over\Gamma(\delta+1)}\Bigr]\,e^{kz}\,dz,\eqno({\rm A}.2)$$
identifies with the one-sided stable L\'evy law $L^{^C}_{\delta,1}(k)$ with $C={t_0^{-\delta}/\Gamma(\delta+1)}$.
We have [4],[26]:
$$\mu_{\rm KWW}(k)={1\over\pi}\,t_0\int_0^\infty\exp\Bigl[-{x^\delta\cos\delta\pi\over\Gamma(\delta+1)}\Bigr]\,\sin\Bigl[{x^\delta\sin\delta\pi\over\Gamma(\delta+1)}\Bigr]\,e^{-kt_0x}\,dx,\eqno({\rm A}.3)$$
hence the associated relaxation times distribution:
$$P_{\rm KWW}(\tau)={1\over\pi}\,{t_0\over\tau^2}\int_0^\infty\exp\Bigl[-{x^\delta\cos\delta\pi\over\Gamma(\delta+1)}\Bigr]\,\sin\Bigl[{x^\delta\sin\delta\pi\over\Gamma(\delta+1)}\Bigr]\,e^{-{xt_0/\tau}}\,dx.\eqno({\rm A}.4)$$

Using a Tauberian theorem [4], it is easy to deduce the behaviour of $P_{\rm KWW}(\tau)$ for $\tau\ll t_0$ from the small-$x$ behaviour of the function $\exp[-{x^\delta\cos\delta\pi\over\Gamma(\delta+1)}]\,\sin[{x^\delta\sin\delta\pi\over\Gamma(\delta+1)}]$.\ As a result, we get:
$$P_{\rm KWW}(\tau)\simeq{\sin\delta\pi\over\pi}\,{1\over t_0}\,\Bigl({\tau\over t_0}\Bigr)^{\delta-1},\qquad\tau\ll t_0.\eqno({\rm A}.5)$$
This result is fully consistent with the form taken by $P_\gamma(\tau)$ (i.e.\ the distribution of relaxation times associated with the Mittag-Leffler function $E_\delta[-{(t/t_0)}^\delta]$) in the same range of values  of $\tau$ (see Eq.\ (3.25)).\ Formulas (3.25) and (A.5) display an integrable divergence of the distribution of small relaxation times, which implies a profusion of short time scales in both the Mittag-Leffler function and the associated stretched exponential [24].\
\bigskip
\noindent
{\bf A.2.\ The relaxation times distribution associated with the algebraically decaying function ${{({t/ t_0})}^{-\delta}\over\Gamma(1-\delta)}$ ($0<\delta<1$)}

\noindent
In the same range of values of $\delta$, consider the algebraic function ${{({t/ t_0})}^{-\delta}\over\Gamma(1-\delta)}$  which pictures the long time behaviour of the Mittag-Leffler function $E_\delta[-{(t/t_0)}^\delta]$.\ Interestingly, this algebraic function can be written as:
$${1\over\Gamma(1-\delta)}\,{\Bigl({t\over t_0}\Bigr)}^{-\delta}={\sin\delta\pi\over\pi}\int_0^\infty{1\over\tau}\,{\Bigl({\tau\over t_0}\Bigr)}^{-\delta}\,e^{-{t/\tau}}\,d\tau.\eqno({\rm A}.6)$$
Eq.\ (A.6) displays the fact that the associated relaxation times distribution is:
$${\sin\delta\pi\over\pi}\,{1\over\tau}\,{\Bigl({\tau\over t_0}\Bigr)}^{-\delta},\eqno({\rm A}.7)$$
a result consistent with the form taken by the function $P_\gamma(\tau)$ as given by Eq.~(3.23) for large relaxation times (i.e., for $\tau\gg t_0$).
\vfill
\break
\noindent
{\smalltitle Appendix B}
\bigskip
\noindent
{\bf B.1.\ The relation between $\hat\gamma(z)$ and $\hat\phi(z)$}

\noindent
Let us here come back in more detail to the relation (3.14) between $\hat\gamma(z)$ and $\hat\phi(z)$, rewritten below for clarity:
$$\hat\gamma(z)={\gamma(t=0)\over z+\hat\phi(z)}\ccomma\eqno({\rm B}.1)$$
together with the generalized Langevin equation (1.1), rewritten in Laplace transforms notations for convenience:
$$z\hat v(z)-v(t=0)+\hat\gamma(z)\hat v(z)={1\over m}\hat F_1(z),\eqno({\rm B}.2)$$
and the Langevin-like equation (3.10) for $F_1(t)$, also rewritten in Laplace transforms notations:
$$z\hat F_1(z)-F_1(t=0)+\hat\phi(z)\hat F_1(z)=\hat F_2(z).\eqno({\rm B}.3)$$

Following Mori's procedure [11], we write the particle's velocity as a sum of two orthogonal components, namely, its projective and vertical components with respect to the $v$ axis (the symbol $v$ denotes the value of $v(t)$ at time $t=0$):
$$v(t)={\langle v(t)v\rangle\over\langle v^2\rangle}\,v+\int_0^t{\langle v(t')v\rangle\over\langle v^2\rangle}{1\over m}f_1(t-t')\,dt',\eqno({\rm B}.4)$$
or, in Laplace transforms notations:
$$\hat v(z)={\int_0^\infty\langle v(t)v\rangle e^{-zt}\,dt\over\langle v^2\rangle}\,v+{\int_0^\infty\langle v(t)v\rangle e^{-zt}\,dt\over\langle v^2\rangle}{1\over m}\hat f_1(z).\eqno({\rm B}.5)$$
Identifying Eqs.\ (B.2) and (B.5), gives:
$${1\over z+\hat\gamma(z)}={1\over\langle v^2\rangle}\int_0^\infty\langle v(t)v\rangle e^{-zt}\,dt,\qquad \hat f_1(z)=\hat F_1(z),\eqno({\rm B}.6)$$
i.e.\  the first fluctuation-dissipation theorem.\ In the same way, still following Mori's procedure, we write the Langevin force as a sum of its projective and vertical components with respect to the $F_1$ axis:
$$F_1(t)={\langle F_1(t)F_1\rangle\over\langle F_1^2\rangle}\,F_1+\int_0^t{\langle F_1(t')F_1\rangle\over\langle F_1^2\rangle}{1\over m}f_2(t-t')\,dt',\eqno({\rm B}.7)$$
or, in Laplace transforms notations:
$$\hat F_1(z)={\int_0^\infty\langle F_1(t)F_1\rangle e^{-zt}\,dt\over\langle F_1^2\rangle}\,F_1+{\int_0^\infty\langle F_1(t)F_1\rangle e^{-zt}\,dt\over\langle F_1^2\rangle}{1\over m}\hat f_2(z).\eqno({\rm B}.8)$$
Identifying Eqs.\ (B.3) and (B.8), gives:
$${1\over z+\hat\phi(z)}={1\over \langle F_1^2\rangle}\int_0^\infty\langle F_1(t)F_1\rangle e^{-zt}\,dt,\qquad\hat f_2(z)=\hat F_2(z)\eqno({\rm B}.9)$$

Now, the second fluctuation-dissipation theorem reads:
$$\langle F_1(t)F_1\rangle=mk_BT\gamma(t),\eqno({\rm B}.10)$$
or,  in Laplace transforms notations:
$$\int_0^\infty\langle F_1(t)F_1\rangle e^{-zt}\,dt=mk_BT\hat\gamma(z).\eqno({\rm B}.11)$$
Comparing Eqs.\ (B.9) and (B.11), gives:
$$\hat\gamma(z)={\langle F_1^2\rangle\over mk_BT}{1\over z+\hat\phi(z)}\cdot\eqno({\rm B}.12)$$
Then, making use of formula (B.10) at time $t=0$, we recover formula (B.1).
\bigskip
\noindent
{\bf B.2.\ Derivation of the noise cut-off function}

\noindent
Interestingly, in the framework of the Mori-Zwanzig formalism, the noise cut-off function is not introduced a priori for convenience purposes, but can be consistently derived once the second-order memory kernel $\phi(t)$ is known.\

Applying standard harmonic analysis to Eq.\ (3.10), we get, introducing the Fourier-Laplace transform $\tilde\phi(\omega)$ of $\phi(t)$:
$$\langle{|F_1(\omega)|}^2\rangle={\langle{|F_2(\omega)|}^2\rangle\over{\bigl|-i\omega+\tilde\phi(\omega)\bigr|}^2}\cdot\eqno({\rm B}.13)$$
On the other hand, the second-fluctuation theorem (1.2) reads, using Fourier transforms:
$$\langle{|F_1(\omega)|}^2\rangle=2mk_BT\RE\tilde\gamma(\omega).\eqno({\rm B}.14)$$
Since we have:
$$\tilde\gamma(\omega)={\gamma_\delta\over t_0^\delta}{1\over -i\omega+\tilde\phi(\omega)}\ccomma\eqno({\rm B}.15)$$
Eq.\ (B.14) reads:
$$\langle{|F_1(\omega)|}^2\rangle=2mk_BT{\gamma_\delta\over t_0^\delta}\,{\RE\tilde\phi(\omega)\over{\bigl|-i\omega+\tilde\phi(\omega)\bigr|}^2}\cdot\eqno({\rm B}.16)$$
From formulas (B.13) and (B.16), we deduce the spectral density of the second-order random force:
$$\langle{|F_2(\omega)|}^2\rangle=2mk_BT{\gamma_\delta\over t_0^\delta}\RE\tilde\phi(\omega),\eqno({\rm B}.17)$$
that is, making use of the identity $\tilde\phi(\omega)=\hat\phi(z=-i\omega)$, together with the expression (3.15) for $\hat\phi(z)$:
$$\langle{|F_2(\omega)|}^2\rangle=2mk_BT\gamma_\delta\sin({\delta\pi/2})t_0^{-2\delta}{|\omega|}^{1-\delta}.\eqno({\rm B}.18)$$

Let us now come back to the spectral density $\langle{|F_1(\omega)|}^2\rangle$ as given by Eq.\ (B.16).\ We have:
$$\langle{|F_1(\omega)|}^2\rangle=2mk_BT{\gamma_\delta\over t_0^\delta}{\RE\tilde\phi(\omega)\over\omega^2{(\omega t_0)}^{-2\delta}}{1\over 1+2\cos({\delta\pi/2}){(|\omega|t_0)}^\delta+{(|\omega|t_0)}^{2\delta}}\ccomma\eqno({\rm B}.19)$$
an expression, which, once the expression for $\RE\tilde\phi(\omega)$ is made explicit, identifies with formula (3.33) of the main text, that is:
$$\langle{|F_1(\omega)|}^2\rangle=2mk_BT\gamma_\delta\sin({\delta\pi/2}){|\omega|}^{\delta-1}\,f_c(|\omega|t_0).\eqno({\rm B}.20)$$
\ The shorter time scale $t_0$ intervenes only in the cut-off function.\ This latter verifies:
$$f_c(|\omega| t_0)={\omega^2{(|\omega|t_0)}^{-2\delta}\over{\bigl|-i\omega+\tilde\phi(\omega)\bigr|}^2}\cdot\eqno({\rm B}.21)$$
\vfill
\break
\parindent=0pt
{\smalltitle References}
\bigskip
\baselineskip=12pt
\frenchspacing
\hfuzz=4pt

[1] N.\ Pottier, Physica A 317 (2003) 371.

[2] S.C.\ Kou, X.S.\ Xie, Phys.\ Rev.\ Lett.\ 93 (2004) 180603.

[3] A.D.\ Vi\~nales, M.A.\ Desp\'osito, Phys.\ Rev.\ E 75 (2007) 042102.

[4] D.V.\ Widder, The Laplace Transform, Princeton University Press, 1941.

[5] M.N.\ Berberan-Santos, E.N.\ Bodunov, B.\ Valeur, Chem.\ Phys.\ 315 (2005) 171.

[6] A.\ Erd\'elyi et al., Higher Transcendental Functions, Vol.\ 3, McGraw Hill, New 

York, 1955.

[7] I.\ Podlubny, Fractional Differential Equations, Academic Press, San Diego, 1999.

[8] F.\ Mainardi, R.\ Gorenflo, J.\ Comput.\ Appl.\ Math.\ 118 (2000) 283.

[9] E.\ Lutz, Phys.\ Rev.\ E 64 (2001) 051106.

[10] W.T.\ Coffey, Y.P.\ Kalmykov, S.V.\ Titov, Adv.\ Chem.\ Phys., Vol.\ 133 Part B, 285, Wiley, New York, 2006.

[11] H.\ Mori, Prog.\ Theor.\ Phys.\ 33 (1965) 423; 

H.\ Mori, Prog.\ Theor.\ Phys.\ 34 (1965) 399.

[12] R.\ Zwanzig, Nonequilibrium statistical mechanics, Oxford University Press, Oxford, 2001.

[13] F.\ Mainardi, M.\ Raberto, R.\ Gorenflo, E. Scalas, Physica A 287 (2000) 468.

[14] R.\ Metzler, J.\ Klafter, J.\ Non-Cryst.\ Solids 305 (2002) 81.

[15] M.N.\ Berberan-Santos, J.\ Math.\ Chem.\ 38 (2005) 629.

[16] K.S.\ Cole, R.H.\ Cole, J.\ Chem.\ Phys.\ 9 (1941) 341.

[17] U.\ Weiss, Quantum Dissipative Systems, third ed., World Scientific, Singapore, 2008.

[18] R.\ Kupferman, J.\ Stat.\ Phys.\ 114 (2004) 291.

[19] R.\ Zwanzig, J.\ Stat.\ Phys.\ 9 (1973) 215.

[20] R.\ Metzler, J.\ Klafter, J.\ Phys.\ Chem.\ B 104 (2000), 3851.

[21] E.\ Barkai, R.J.\ Silbey, J.\ Phys.\ Chem.\ B 104 (2000), 3866.

[22] S.\ Burov, E.\ Barkai, Phys.\ Rev.\ Lett.\ 100 (2008) 070601; 

S.\ Burov, E.\ Barkai, Phys.\ Rev.\ E 78  (2008) 031112.

[23] B.J.\ Berne, R.\ Pecora, Dynamic Light Scattering, Wiley-Interscience, New York, 1976 (Reprinted, Dover Publications, New York, 2000).

[24] J.-P.\ Bouchaud, Anomalous relaxation in complex systems: from stretched to compressed exponentials, in: G.\ Radons, R.\ Klages, and I.M.\ Sokolov (Eds.), Anomalous Transport. Foundations and Applications, Wiley-VCH,Weinheim, 2008, pp. 327--345.

[25] J.-P.\ Bouchaud, A.\ Georges, Phys.\ Rep.\ 195 (1990) 127.

[26] E.W.\ Montroll, J.T.\ Bendler, J.\ Stat.\ Phys.\ 34 (1984) 129.

\bye